\documentclass[aps,showpacs]{revtex4}

\usepackage{amssymb}
\usepackage{graphicx}
\usepackage{bm}
\usepackage{amsmath}

\begin{document}

\title{Fast Impurity Solver for Dynamical Mean Field Theory
based on second order perturbation around the
atomic limit}
\author {Jia Ning Zhuang, Qing Mei Liu, Zhong Fang, Xi Dai$^*$}


\begin{abstract}
An impurity solver  for the dynamical mean field (DMFT) study of the
Mott insulators is proposed, which is based on the second order
perturbation of the hybridization function. After carefully
benchmarking it with Quantum Monte Carlo  results on the anti-ferromagnetic phase of the
Hubbard model, we conclude that this impurity solver can capture the
main physical features in the strong coupling regime and
can be a very useful tool for the LDA+DMFT studies of the Mott
insulators with long range order.
\end{abstract}

 \maketitle

\section{Introduction}

Mott insulators are insulators which are caused by
electron-electron interaction rather than the crystal  structure. In
sufficient low temperature, long range spin or orbital order will be
established to get rid of the remaining entropy induced by the
unfrozen spin or orbital degree of freedoms. Various types of
interesting long range order may appear in different systems,
i.e. the anti-ferromagnetic (AF) order, the non-collinear spin order
as well as the orbital ordering states. The fruitfulness of the long
range orders in the Mott insulator phases is a consequence of the
detailed balance among electron kinetic energy, interaction energy
and the spin orbital coupling in the solid. Therefore the \textit{ab initio} study of
the Mott insulators is very crucial to achieve quantitative
understanding about these materials.

While the first principle calculations for the Mott insulators based on local density approximation (LDA)
always encounter problems generated by the strong correlation effect among those partially occupied
localized orbitals in these materials. For example, the LDA calculation incorrectly predict  two typical Mott
Insulators $NiO$ and $FeO$ to be metallic. This is due to the non-interacting form of
the LDA type wave function can not capture the electron
correlation effect, especially the strong on-site correlation effect
in the relatively local orbitals such as $3d$ or $4f$ orbitals. Although LDA underestimates the strong
correlation effects in these materials, it still describes quite well the long range Coulomb
interaction and those less correlated \textit{s p} bands.
Therefore a very efficient way to solve this problem is to implement
the LDA with some many body techniques which can treat the
correlation effect better. The first example of such scheme is the LDA+U method \cite{LDA+U}, in which the Hartree-Fock mean field treatment is adopted
to capture the orbital physics. The LDA+U method can correctly
describe many Mott insulators, i.e. the $YVO_3$ and $LaVO_3$ \cite{FZ}. While since
the wave function in LDA+U is still the single slater
determinant, it can only capture the static orbital or spin
correlation but not the dynamical correlation effect.  Another attempt is the
LDA+Gutzwiller method proposed very recently by two of the
authors, where the Gutzwiller variational approach \cite{Gutzwiller_old} \cite{He3}  is used to take into account the
correlation effect (LDA+G). LDA+Gutzwiller is superior to LDA+U in
the sense that the dynamical correlation effect can be taken into
account by the multi-configuration nature of the Gutzwiller trial
wave function. However, LDA+G is still very difficult to be used in
the finite temperate problems because it is a variational
method.

In the past twenty years, the dynamical mean field theory (DMFT) \cite{DMFT} has been quickly developed to be a powerful method to solve
the strongly correlated models on the lattice. DMFT maps the lattice
models to the corresponding quantum impurity models with the
environment to be determined self-consistently by the DFMT self consistent
equations. DMFT keeps the full local dynamics induced by the local
interaction and has been successfully applied to various of
correlation systems, such as the Mott transition in Hubbard model
\cite{mit1} \cite{mit2}, the pseudo gap behavior in high Tc
cuperates \cite{pseudo_gap} as well as the heavy fermion system
\cite{qimiao} \cite{qimiao2}.  Combined with LDA, the LDA+DMFT has
quickly become a very useful numerical tool for the first principle
studies of strongly correlated materials, which has been successfully applied
to many strongly correlated materials.

The nuclei of DMFT is to  solve the Anderson impurity model
introduced by the DMFT self consistent procedure. The solver could
be either numerical and analytical. The most popular two numerical
solvers are exact diagonalization \cite{ED} \cite{exactdiag} method
and Hirsh-Fye quantum Monte Carlo \cite{QMC} \cite{QMC1}
\cite{Hirsh-Fye}. Besides, the  continuous-time quantum Monte Carlo (CTQMC) solver
\cite{CTQMC} \cite{weak_ctqmc} \cite{Pu_Gabi} \cite{LaOFeAs_Gabi}
has also been developed recently and greatly boosted the progress in this field .
However, the numerical methods
are often very time-consuming and for large
systems it is better to use some analytical solver which is much faster and can capture
the basic physics as well. The analytical solvers include Hubbard-I
approximation \cite{Savrasov-Hub1} \cite{Savrasov-Hub1_2}, the
equation of motion (EOM) method \cite{eom}, the iterative
perturbation theory \cite{IPT1} \cite{IPT2}, the fluctuation
exchange approximation(FLEX) \cite{FLEX} and the non-crossing
approximation \cite{NCA} \cite{NCA2}. Among them the Hubbard-I
solver, which simply takes the atomic self energy, is very
convenient and easy to be implemented in the existing \textit{ab initio}
codes. Since it can capture the atomic multiplet effect (Hubbard
bands) in a very efficient  way, the Hubbard-I solver has been widely
used to study the paramagnetic Mott insulators. While the Hubbard-I
solver can not be directly applied to the Mott insulator states with
long range order, simply because the atomic self energy does not contain
any information about the long range order in the system. The way to
solve this problem is to go beyond the atomic approximation and
include the leading order corrections from the heat bath as well.

A second order strong coupling expansion solver has been developed
by one of the authors in reference \cite{atom3}. Although the long
range ordering temperature can be correctly calculated using this
solver, it also suffers from the following facts. i) Because the
strong coupling expansion is based on the atomic limit without long
range order, it is not valid to treat the systems with large order
parameter, i.e. the AF phase in low temperature with the
fully developed AF order. ii)Due to the multiple summation over the
configuration space, the calculation is still quite heavy for the
multi-orbial system especially the \textit{f}-electron system.

In order to overcome the above two problems, in this paper we
propose a new scheme to include leading order corrections from the
heat bath. The idea is quite simple: we use the second order
perturbation theory to correct the energy level of each atomic
eigenstates. Then we construct the ``atomic" Green's function  and the corresponding self energy
on a single site based on the re-normalized atomic eigen-energy.
 With the above procedure, we
can obtain the main contribution of the heat bath to the self energy
in the Mott insulator case, which is the re-normalization of the eigen-energy
of the atomic configurations.  Because only the eigen-energy not the occupation
of the atomic configurations   is treated by second order perturbation, the present
method can be applied to very low temperature with quite large order parameters.
With the present impurity solver the static inter-site
spin or orbital correlation, such as the super-exchange process, can
be taken into account. Compared with the previous atomic expansion solver, this
solver does not have multiple summations over the atomic
configurations, which make it very fast and is more suitable for the
study of the \textit{f}-electron systems.

The paper is organized as the following. We will present the detailed formalism of the impurity solver
in section II. After that in section III, we will benchmark it on both paramagnetic and anti-ferromagnetic Mott
insulator phases in the Hubbard model. And finally we will carefully discuss the results and make the conclusion.

\section{Derivations}
\subsection{General Formalism}
The crucial part of the dynamical mean field theory is to solve the
quantum impurity model with the following Hamiltonian:
\begin{eqnarray}
\nonumber H_{imp} &=&\sum_{l} \epsilon_{l}
c_{l}^{\dagger}c_{l}+\sum_{lm}(V_{lm}f_m^{\dagger}c_{l}+h.c.)\\
\nonumber &+&\sum_{mm'}t_{mm'}f_m^{\dagger}f_{m'}
+\frac{1}{2}\sum_{mnpq}U_{mnpq}f^{\dagger}_mf^{\dagger}_nf_qf_p\\
\label{Himp}
\end{eqnarray}
The above expression is a general form of multi-orbital Anderson
impurity model with arbitrary local interactions as well as the
heat bath introduced by DMFT self consistent procedure.
In this Hamiltonian, $m$, $m'$ and $n$, $p$, $q$
are the combined spin and orbital indices, while $l$ denotes the energy levels in
the heat bath.

In the LDA+DMFT study, we often need a fast impurity solver which
can capture the most important features of the impurity problem. For
the Mott insulators, such a fast solver can be constructed by doing
perturbation around the atomic limit. In the present paper, we apply
the ``contour integral" technique to obtain the electron Green's
function \cite{NCA}. The retarded form of the electron Green's
function on the impurity site can be written as
\begin{eqnarray}
\nonumber G_{f}^{m
m'}(\omega+\mathrm{i}\eta)&=\frac{1}{\mathcal{Z}}\oint_c\frac{\mathrm{d}z}{2\pi\mathrm{i}}\mathrm{e}^{-\beta
z}\sum\limits_{\alpha\beta}(F^m)_{\alpha\beta}(F^{m'\dagger})_{\beta\alpha}\\
&\times
P_{\beta}(z)\Pi_{\beta,\alpha}(z,z+\omega+\mathrm{i}\eta)P_\alpha(z+\omega+\mathrm{i}\eta)\label{Gimp}\\
&\mathcal{Z}=\oint_c
\frac{\mathrm{d}z}{2\pi\mathrm{i}}\mathrm{e}^{-\beta
z}\sum\limits_{\alpha}P_{\alpha}(z)
\end{eqnarray}
In the above equations, $c$ is a contour in the complex plane, oriented
counterclockwise, and surrounding all singularities of the resolvent
$(z-H_{loc})^{-1}$, where $H_{loc}$ is the local part of $H_{imp}$;
$\alpha$ denotes the $\alpha$th eigen state of $H_{loc}$ with the eigen value
of $E_{\alpha}$; $(F^m)_{\alpha\beta}$ is the matrix element of
local fermion operator $f_m$;
$\Pi_{\beta,\alpha}(z,z+\omega+\mathrm{i}\eta)$ is a proper vertex
function to be discussed below; $\mathcal{Z}$
is the partition function of the impurity problem and $P_{\alpha}(z)$ can be viewed as the
"propagator" of the atomic eigen states. In the absence of hybridization
takes the simple form of $P_{\alpha,0}(z)=1/(z-E_{\alpha})$. When
taking the hybridization part into account,
$P_{\alpha}(z)$ can be expressed as
\begin{eqnarray}
P_{\alpha}(z)=1/(z-E_{\alpha}-\Sigma_{\alpha}(z))\label{localspectrum}
\end{eqnarray}
where $\Sigma_{\alpha}(z)$ can be viewed as the self-energy of the
$\alpha$th local configuration that comes from the hybridization
between the impurity site and the bath.

\begin{figure}[tbp]
\begin{center}
\includegraphics[width=12cm,angle=0,clip=]{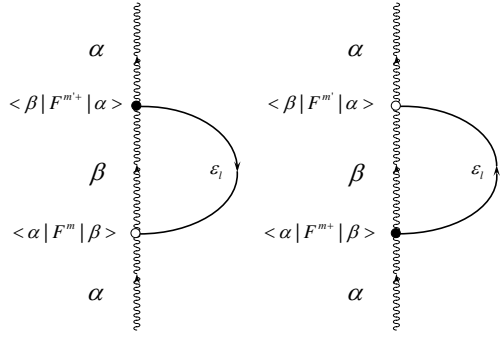}
\end{center}
\caption{The diagrams of the lowest order contribution to the
self energy of local atomic states. The break lines denote the
propagators of atomic states, the continuous lines the
propagator of electrons in the heat bath, and each vertex corresponds to
an element of a local Fermion operator with the coupling strength
$V$.}\label{sepic}
\end{figure}

In order to evaluate $\Sigma_{\alpha}(z)$, we use the method
heuristic from the techniques in the large-N expansion. The self
energy can be expressed in a diagrammatic way as it is introduced
in detail in the review \cite{LargeN}. In this paper, we focus on the Mott insulator system,
which only requires  the lowest  order diagram as shown in Fig \ref{sepic}.

The expression of self-energy corresponding to the above diagram is:
\begin{eqnarray}
\nonumber\Sigma_{\alpha}(z)=\sum\limits_{\beta}\sum\limits_l\sum\limits_{mm'}\frac
{(F^m)_{\alpha\beta}(F^{m'\dagger})_{\beta\alpha}V_{lm'}^*V_{lm}}
{z-E_{\beta}+\epsilon_l}f(\epsilon_l) \\
\nonumber+\sum\limits_{\beta}\sum\limits_l\sum\limits_{mm'}\frac
{(F^{m\dagger})_{\alpha\beta}(F^{m'})_{\beta\alpha}V_{lm}^*V_{lm'}}
{z-E_{\beta}-\epsilon_l}f(-\epsilon_l) \\
\nonumber=(-\frac{1}{\pi})\sum\limits_\beta\sum\limits_{mm'}
\big[(F^m)_{\alpha\beta}
(F^{m'\dagger})_{\beta\alpha}\int\frac{\Delta''_{m'm}(-\epsilon)f(-\epsilon)}{z
-E_\beta-\epsilon}\mathrm{d}\epsilon\\
+(F^{m\dagger})_{\alpha\beta}(F^{m'})_{\beta\alpha}\int\frac{\Delta''_{mm'}
(\epsilon)f(-\epsilon)} {z-E_\beta-\epsilon}\mathrm{d}\epsilon\big] \label{localse}
\end{eqnarray}
where $f(\epsilon_l)$ is the fermi distribution function, while
\begin{eqnarray}
\Delta_{mm'}(\omega+\mathrm{i}\eta)=\sum\limits_l\frac{V_{lm}^*V_{lm'}}{\omega+\mathrm{i}\eta-\epsilon_l}
\end{eqnarray}
is called the hybridization function in the dynamical mean field
theory and its imaginary part is denoted as
\begin{eqnarray}
\Delta''_{mm'}(\epsilon)=\mathrm{Im}[\Delta_{mm'}(\epsilon+\mathrm{i}\eta)]
\end{eqnarray}
in equation (\ref{localse}).

Afterwards, for the sake of a simplest self-consistent approximation
that provides a soluble set of equations, the vertex function in
Eqn.(\ref{Gimp}) could be set
\begin{eqnarray}
\Pi_{\alpha,\alpha'}(z,z')\equiv \mathbf{1}
\end{eqnarray}
and we now get the final expression of $G_{f}^{m
m'}(\omega+\mathrm{i}\eta)$ in Eqn.(\ref{Gimp}).

The local electrons' quasi-particle spectrum functions are the
direct result of the retarded form of the Green's functions.
For a given Hamiltonian as
Eqn.(\ref{Himp}), one may use the equations above to evaluate the
Green's function. Implemented with the DMFT self-consistent
condition described in the review \cite{DMFT}, this method can be
used as an impurity solver in the DMFT study of the multi-orbital
systems with complicated local interactions.

\subsection{Single Pole Approximation}
The method shown above works well for Mott insulator phase, and a few results will be shown
in the next section. However, here we introduce a further
approximation which makes the impurity solver much faster and more
convenient. The main idea of this approximation can be interpreted
as the following.  For a Mott insulator solution  in DMFT  with long range order ,
the most important effect of the heat bath is to modify the effective energy level
of the atomic configurations. Therefore in the present paper, we only consider the
effect of heat bath in  the renormalization of the atomic levels, but neglect the possible broadening
 of the atomic levels.

With the above consideration,  Eqn.(\ref{localspectrum}) can be approximated as
\begin{eqnarray}
\nonumber P_{\alpha}(z)&=&1/(z-E_{\alpha}-\Sigma_{\alpha}(z))\\
\nonumber
&\approx&1/(z-E_{\alpha}-\mathrm{Re}[\Sigma_{\alpha}(E_{\alpha})])\\
&\equiv&1/(z-\tilde{E}_{\alpha})
\end{eqnarray}
, and $\tilde{E}_{\alpha}\equiv
E_{\alpha}+\mathrm{Re}[\Sigma_{\alpha}(E_{\alpha})]$ is the
renormalized atomic level containing the second order correction of the hybridization function.
Because the Green's function
$P_{\alpha}(z)$ contains only one pole at
$\tilde{E}_\alpha$, as the same form in unperturbed case, we call it ``single pole approximation".
 Thus the electron Green's function can be written in a similar form as the atomic Green's function
\begin{eqnarray}
G_f^{mm'}(\omega+\mathrm{i}\eta)&=\frac{1}{Z_f}\sum\limits_{\alpha\beta}(F^m)_{\beta\alpha}
(F^{m'\dagger})_{\alpha\beta}\\
&\frac{1}{\omega+\mathrm{i}\eta+\tilde{E}_\beta-\tilde{E}_\alpha}
\big[\mathrm{e}^{-\beta\tilde{E}_\beta}+\mathrm{e}^{-\beta\tilde{E}_\alpha}\big]\label{gf2}
\end{eqnarray}
where the partition function is $Z_f=\sum\limits_\alpha\mathrm{e}^{-
\beta\tilde{E}_\alpha}$. Correspondingly, the self energy that comes
from local two-particle interactions becomes
\begin{eqnarray}
\Sigma(\omega+\mathrm{i}\eta)=\omega+\mathrm{i}\eta+\mu-\hat{t}-G_f^{-1}\label{se2}
\end{eqnarray}
This self energy can be implemented in the DMFT self consistent loop and we obtained
a fast impurity solver for the Mott insulators with the long range order fully developed. With
the current approximation, the DMFT here is equivalent to a generalized static mean field
approximation using the local many-body bases. The spontaneous symmetry breaking can be
well described by the difference in energy and occupation for atomic configurations which can be
connected by symmetry, i.e. the singly occupied spin up and down states in the one-band Hubbard
model. Therefore the super-exchange process can be thus captured through the self consistently
determination of the  hybridization  function in the DMFT loop.

\section{Results and discussion}

\subsection{Single band, Paramagnetic phase}

\begin{figure}[tbp]
\begin{center}
\includegraphics[width=8cm,angle=0,clip=]{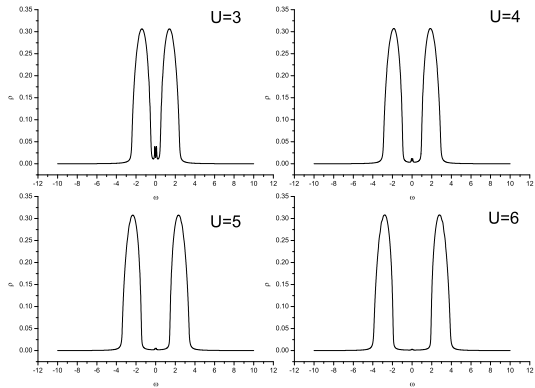}
\end{center}
\caption{The spectral function of single band Hubbard model on the
Bethe lattice at half filling for various values of $U$, the energy unit is chosen as D=2t=1.}
\label{singlepara}
\end{figure}

The spectral function of one-band Hubbard model with various $U$ at
half filling is plotted in Fig \ref{singlepara}. It works on the
Bethe lattice with the semicircular density of states

\begin{eqnarray}
D(\epsilon)=\frac{1}{2\pi t^2}\sqrt{4t^2-\epsilon^2},\quad
|\epsilon|<2t
\end{eqnarray}
where $t=0.5$ as usual. 
Here we do not use the
``single pole approximation", which means Eqn.(\ref{localse}) is
implemented. As a perturbation method around atomic limit, we echo
the fact that this kind of solver is limited to integer filling and
large $U$ \cite{atom3}, that is, the present solver is just good to
study in the regime of Mott insulating state.

\begin{figure}[tbp]
\begin{center}
\includegraphics[width=8cm,angle=0,clip=]{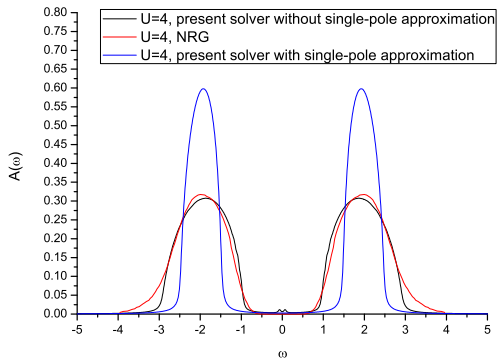}
\end{center}
\caption{The spectral function for $U=4$ on the Bethe lattice
compared with DMFT+NRG, the energy unit is chosen as D=2t=1.} \label{compareNRG}
\end{figure}

As a comparison, we also exhibit the result of present solver with and without single pole 
approximation compared with that
of DMFT+NRG, see Fig \ref{compareNRG}. 
It shows that the result of present solver at large $U$ regime agrees
well with that of NRG, and the single pole approximation captures the main features
of the Hubbard bands.

\subsection{Single band, Antiferromagnetic-Paramagnetic Phase Transition}

\begin{figure}[tbp]
\begin{center}
\includegraphics[width=8cm,angle=0,clip=]{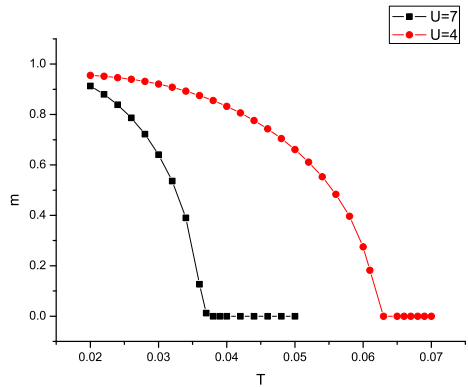}
\end{center}
\caption{The sublattice magnetisation $m$ versus temperature $T$ for
$U=4,7$,  the energy unit is chosen as D=2t=1.} \label{orderparameter}
\end{figure}

\begin{figure}[tbp]
\begin{center}
\includegraphics[width=14cm,angle=0,clip=]{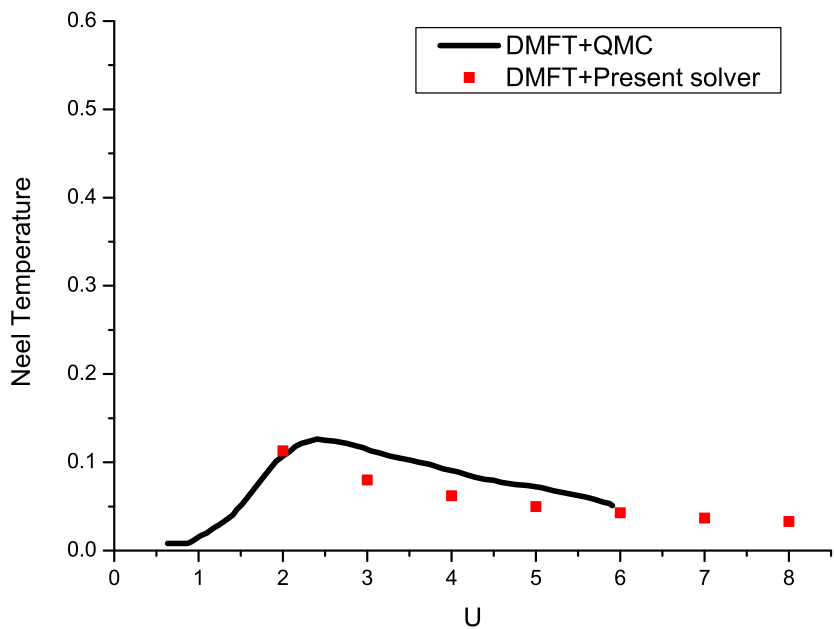}
\end{center}
\caption{The comparison of the N\'eel temperature obtained by
DMFT+QMC (thick line) and present solver (circle dots) on the Bethe
lattice at half filling. The half band width is used as the energy unit by defining 
$D=2t=1$.}\label{Tcsemi}
\end{figure}

In this section, we study the antiferromagnetic(AF)-paramagnetic phase
transition in single band Hubbard model at half filling. We focus on
the AF order parameter $m=\langle
n_{\uparrow}-n_{\downarrow}\rangle$ on a given sublattice the
N\'eel temperature. Henceforth we use the single pole approximation,
with the self energy in Eqn.(\ref{se2}). First we study the single band Hubbard model on the Bethe
lattice with the semicircular density of state.
Fig.\ref{orderparameter} shows the magnetisation $m$ on one
sublattice versus temperature $T$ for $U=4$ and $U=7$. The curves show 
that the phase transition is of second order, and
the critical temperatures are about $T_c=0.061$ for $U=4$ and
$T_c=0.037$ for $U=7$ respectively. Likewise, we get $T_c$ under
different $U$ and plot them in Fig.\ref{Tcsemi}. This result is
comparable to that of DMFT+QMC in large $U$ regime, as it is
expected.

\begin{figure}[tbp]
\begin{center}
\includegraphics[width=8cm,angle=0,clip=]{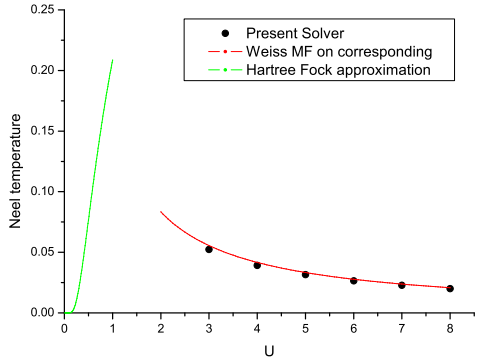}
\end{center}
\caption{The comparison of the N\'eel temperature obtained by Weiss
MF and present solver on the 3D cubic lattice at half filling. For cubic 
lattice the unit energy is chosen as the
half band width $D=6t=1$.}\label{Tccubic}
\end{figure}

We also apply this method to estimate the N\'eel temperature on the
3D isotropic simple cubic lattice. As the dispersion relation is $
E(\mathbf{k})=2t(\mathrm{cos}k_x+\mathrm{cos}k_y+\mathrm{cos}k_z)$ ,
we choose $t=1/6$ to ensure that the half band width equals to $1$.
It is known that, in the large $U$ limit, on the 3D cubic lattice,
the Hubbard model is equivalent to a Heisenberg model with an
antiferromagnetic coupling $J$, and the relation between the
parameters reads $J=4t^2/U$ \cite{j=4t2/u}. Accordingly, the correct
N\'eel temperature of the model is $T_N=3.83t^2/U$ \cite{correcttn}
while the Weiss mean field gives the result $T_N=6t^2/U$
\cite{mftn}. Remember that we only take the lowest order of
corrections to the local interactions, so the result should be close
to the one of Weiss mean field, see Fig \ref{Tccubic}. We have also
plotted in this diagram the Hartree-Fock Neel temperature for  the same model
\cite{ECM}, which is completely wrong in the large U limit.

\begin{figure}[tbp]
\begin{center}
\includegraphics[width=8cm,angle=0,clip=]{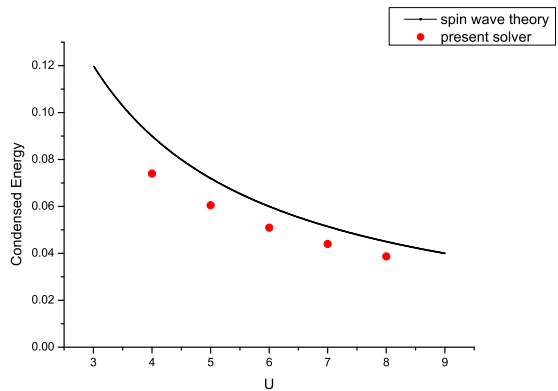}
\end{center}
\caption{The antiferromagnetic condensed energy compared with the
spin-wave theory for Heisenberg model, where the unit energy is
$D=6t=1$.}\label{econdensed}
\end{figure}

It is intuitive to compare our results to that of Heisenberg model,
because the latter has been studied intensively. For example, 
we have compared the antiferromagnetic condensed energy. In
the antiferromagnetic spin-wave theory for Heisenberg model, this
condensed energy is $12.96t^2/U$ on 3D simple cubic lattice
\cite{LZZ}, about twice as much as $T_N$ of Weiss mean field. Here
we have calculated the energy difference between antiferromagnetic
and paramagnetic solutions, at a temperature of $0.002t$ which is
sufficiently low and plotted it in Fig \ref{econdensed} together with the condensation energy
obtained by spin wave theory for the corresponding Heisenberg model. Our results are in good
agreement with the spin wave theory which indicate that our method works very well even in low
temperature.

\subsection{Two band, Antiferromagnetic-Paramagnetic Phase Transition}
We have also studied antiferromagnetic-paramagnetic phase transition
in two-band Hubbard model at half filling, using DMFT with single pole
approximation. Because there are two band on one single site, the
situation becomes much more complicated. Here we only consider the
simplest case. First, we consider two degenerate band, with
semicircular density of states and the band width is set to be $1$.
Second, we neglect the hopping between different orbits. Therefore the Hamiltonian 
can be written  as
\cite{Gutzwiller}
\begin{eqnarray}
\nonumber\hat{H}_{\mathrm{at}}=U\sum\limits_b\hat{n}_{b,\uparrow}\hat{n}_{b,\downarrow}
+U'\sum\limits_{\sigma,\sigma'}\hat{n}_{1,\sigma}\hat{n}_{2,\sigma'}
-J\sum\limits_{\sigma}\hat{n}_{1,\sigma}\hat{n}_{2,\sigma}\\
\label{twobandH}
\end{eqnarray}
There are three adjustable parameters above, which are $U$ the
intra-band Coulomb repulsive energy, $U'$ the inter-band Coulomb
energy, and $J$ the interaction between different bands with the
same spin. For $J>0$, the third term on the right hand side in
Eqn.(\ref{twobandH}) contains the Hund's coupling $-J'S_1^zS_2^z$
since $S^z=1/2(n_\uparrow-n_\downarrow)$. The discussions below are
based on this Hamiltonian.



\begin{figure}[tbp]
\begin{center}
\includegraphics[width=8cm,angle=0,clip=]{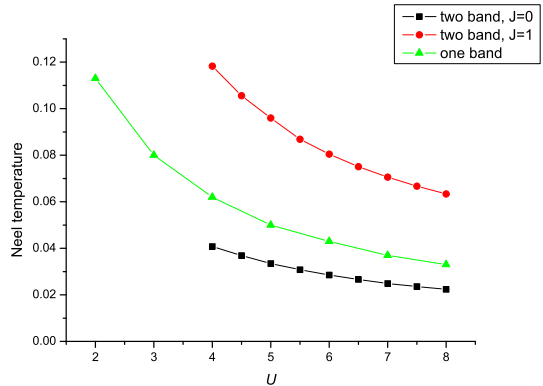}
\end{center}
\caption{Two band model on the Bethe lattice 
under the condition of
$U-U'=2J$
for $J=0$ and $J=1$ 
 respectively, compared with one band result. The energy unit
here for the Bethe lattice is chosen as 
$D=2t=1$.}\label{J}
\end{figure}

Here we further assume
$U-U'=2J$, which is true for systems with the cubic symmetry \cite{U-U'=2J}. 
The N\'eel temperatures versus various $U$ obtained by us  for
$J=0$ and $J=1$ are shown in
Fig.\ref{J} respectively.

The N\'eel temperature of two band Hubbard model with $U=U'$ is
lower than that of one band model with the same $U$. This can be understood
in the way 
that in a two band system at half filling, there is AF order developed in two different orbitals.
But there is no term to lock the relative fluctuation between them, which reduces its tendency to
be anti-ferromagnetic. With non-zero Hund's rule coupling $J$, such relative fluctuation between
AF order in different orbitals will be greatly suppressed, which result in the dramatic increment of the N\'eel
temperature as shown in Fig.\ref{J}.

\section{Conclusions}

This paper presents a new impurity solver based on the perturbation
theory in the hybridization around the atomic limit. For
the sake of speed and convenience, we have introduced a further
``single pole approximation" with which we carefully studied the
antiferromagnetic-paramagnetic phase transition in one band and two
band Hubbard model at half filling. The DMFT with the current impurity solver thus
provide a way to do mean field approximation with the atomic configurations rather
than energy levels for the single particle states. Comparing with the DMFT+QMC
on the single band Hubbard model, the Neel temperature obtained by this new impurity
solver fits quite well with the DMFT+QMC results, which indicate that is can be very useful
in the LDA+DMFT studies for Mott insulators.
 \\

 ACKNOWLEDGEMENT: The authors would thank X. Y. Deng, L.Wang, W. Zhang
 and J. T. Song for their helpful discussions.

\end{document}